\documentclass[aps,prd,twocolumn,superscriptaddress,floatfix,showpacs,showkeys,preprintnumbers]{revtex4}

\usepackage[utf8]{inputenc}

\usepackage[T1]{fontenc}

\usepackage{slashed}

\usepackage{times}

\usepackage{epstopdf}

\usepackage{amssymb,amsfonts,amsmath,amsthm}

\usepackage{dsfont,bbm}

\usepackage{dcolumn}

\usepackage{epsfig}

\usepackage{graphicx}

\usepackage[caption=false]{subfig}

\usepackage{dsfont}

\usepackage{slashed}

\usepackage[active]{srcltx}

\usepackage[usenames]{color}

\newcommand{\up}{{\uparrow}}

\newcommand{\down}{{\downarrow}}

\newcommand{\ide}[1]{\hskip #1 & \hskip -#1}

\begin{document}


\title{Electroproduction of the $N^\ast(1535)$ nucleon resonance in QCD}
\author{I.V.~Anikin}
\affiliation{Institut f\"ur Theoretische Physik, Universit\"at
   Regensburg,D-93040 Regensburg, Germany}
\affiliation{Bogoliubov Laboratory of Theoretical Physics, JINR, 141980 Dubna, Russia}
\author{V.M.~Braun}
\affiliation{Institut f\"ur Theoretische Physik, Universit\"at
   Regensburg,D-93040 Regensburg, Germany}
\author{N.~Offen}
\affiliation{Institut f\"ur Theoretische Physik, Universit\"at
   Regensburg,D-93040 Regensburg, Germany}
\date{\today}
\begin{abstract}

  \vspace*{0.3cm}

\noindent
Following the 12 GeV upgrade, a dedicated experiment is planned with the Hall B CLAS12 detector
at  Jefferson Lab, with the aim to study electroproduction of nucleon resonances at
high photon virtualities up to $Q^2 = 12$~GeV$^2$.
In this work we present a QCD-based approach to the theoretical interpretation of these
upcoming results in the framework of light-cone sum rules that combine perturbative calculations with
dispersion relations and duality.
The form factors are thus expressed in terms of $N^\ast(1535)$ light-front wave functions at small
transverse separations, called distribution amplitudes.
The distribution amplitudes can therefore be determined from the comparison with the
experimental data on form factors and compared to the results of lattice QCD simulations.
The results of the corresponding next-to-leading order calculation are presented and compared with the
existing data. We find that the form factors are dominated by the twist-four distribution amplitudes
that are related to the $P$-wave three-quark wave functions of the $N^\ast(1535)$, i.e. to contributions of orbital
angular momentum.
 \end{abstract}



\maketitle

\date{\today}


\section{Introduction}
\setcounter{equation}{0}


It is generally accepted that studies of baryon form factors at large momentum transfer $Q^2$ give access
to the light-front wave functions at small transverse separations between the constituents,
called hadron distribution amplitudes (DAs), although perturbative QCD factorization
\cite{Chernyak:1977as,Radyushkin:1977gp,Lepage:1979zb} does not seem to be applicable
for realistic $Q^2$ accessible in current or planned experiments.
The problem is that the leading contribution involves two hard gluon exchanges and is suppressed
by the small factor $(\alpha_s/\pi)^2 \sim 0.01$ compared to the ``soft''
(end-point) contributions which are subleading in the power counting in $1/Q^2$ but do not involve small coefficients.
Hence the collinear factorization regime is approached very slowly.
Model calculations suggest that ``soft'' contributions play the dominant role at present energies.
Taking into account soft contributions is challenging because
they involve a nontrivial overlap of nonperturbative wave functions of the initial and the final state
hadrons, and are not factorizable, i.e. cannot be simplified further in terms of simpler quantities.

In this situation the question what exactly do we learn from the studies of form factors is far from trivial.
One existing description is to introduce more complicated, transverse-momentum dependent (TMD) quark distributions,
taking advantage of Sudakov suppression of large transverse separations, following the
technique suggested initially by Li and Sterman~\cite{Li:1992nu} for the pion form factor.
Another approach that we advocate in this work, is to calculate the soft contributions to the form factors as an
expansion in terms of nucleon DAs of increasing twist using dispersion relations and duality. This method is
known as light-cone sum rules (LCSRs)~\cite{LCSR} and provides one with the most direct relation of the
hadron form factors and DAs that is available at present, with no other nonperturbative parameters.

The LCSR approach has been used successfully  for the calculations
of pion electromagnetic and also weak $B$-decay form factors, see
Refs.~\cite{Braun:1999uj,Ball:2004ye,Duplancic:2008ix} for several recent state-of-the-art calculations.
The LCSRs for baryon form factors are more complicated and recent.
The first application was for the nucleon electromagnetic form factors
in Refs.~\cite{Braun:2001tj,Braun:2006hz}.  Several further studies aimed at
finding an optimal nucleon interpolation current~\cite{Lenz:2003tq,Braun:2006hz,Aliev:2007qu,Aliev:2008cs}
and extending this technique to other elastic or transition form factors of interest.
LCSRs for the axial nucleon form factor were presented in \cite{Braun:2006hz,Aliev:2007qu,Wang:2006su},
for the scalar form factor in \cite{Wang:2006su} and tensor form factor in \cite{Erkol:2011iw}.
A generalization to the full baryon octet was considered e.g. in ~\cite{Liu:2009mb}.
Application of the same technique to $N\gamma\Delta$ transitions was suggested
in~\cite{Braun:2005be,Aliev:2007qu} and to pion production at threshold in~\cite{Braun:2006td}.
LCRSs for weak baryon decays $\Lambda_b\to p,\Lambda\ell\nu_\ell$ etc. were studied in
\cite{Huang:2004vf,Wang:2009hra,Aliev:2010uy,Khodjamirian:2011jp}, etc.
In the early work only the leading order (LO) contributions to the coefficient functions in the LCSRs have been taken into account.
The first complete next-to-leading order (NLO) analysis was done for the electromagnetic nucleon form factors
in Ref.~\cite{Anikin:2013aka} and the results appear to be consistent with the constraints on nucleon DAs from lattice
calculations~\cite{Braun:2014wpa}. The picture emerging from these studies sugggests that
the momentum fraction distribution of the valence quarks in the proton is rather broad, with $\sim 40\%$ of the momentum
carried by the $u$-quark that carries proton helicity, and approximately symmetric to the
interchange of the remaining quarks.

Our study is motivated by the dedicated experiment planned with the Hall B CLAS12 detector
at  Jefferson Lab following the 12 GeV upgrade, with the aim to study electroproduction of nucleon resonances
at high photon virtualities up to $Q^2 = 12$~GeV$^2$~\cite{Aznauryan:2012ba}.
The corresponding form factors can be calculated using the LCSR machinery in terms
of the DAs of nucleon resonances.  Turning this relation around, information on the DAs of resonances
can be extracted from the comparison of the LCSR calculations with the experimental data on form factors and compared
to the constraints that can come, eventually, from lattice QCD simulations.
This program was suggested in Ref.~\cite{Braun:2009jy} and an exploratory study was made there
for the particular case of electroproduction of the lowest negative parity $N^*(1535)$ resonance.
In our paper we elaborate on this proposal.
Learning about quark distributions in nucleon resonances
is an exciting possibility, since existing QCD calculations of resonance properties, e.g. on the lattice, rarely go
beyond the mass spectrum.

The case of $N^*(1535)$ is special because the classification and the structure of the light-front wave functions for the states
with opposite parity is almost identical. Hence the LCSRs for the corresponding electroproduction form factors are very similar to the
LCSRs for electromagnetic nucleon form factors. In particular the NLO expressions derived in Ref.~\cite{Anikin:2013aka} can be
overtaken with relatively minor modifications. A detailed analysis of these NLO LCSRs is the main goal of this work.

The presentation is organized as follows.
The electroproduction form factors are introduced and the structure of the corresponding LCSRs is explained in Sect.~2.
Sect.~3 contains a detailed numerical analysis and comparison with the existing experimental data, and Sect.~4 is reserved for a summary and outlook.
A large Appendix~A contains a short review of the three-quark light-front wave functions of
$J^P = \left(\frac{1}{2}\right)^-$ nucleon resonances, their relation to DAs, and also explains our conventions.
Appendix B contains a simple parametrization of the $Q^2$ dependence of the coefficient functions in the LCSRs.

\section{Electroproduction form factors and Light Cone Sum Rules}

The matrix element of the electromagnetic current $j^{\rm em}_\nu$ between spin-1/2
states of opposite parity can be parametrized in terms of two independent
form factors, which can be chosen as
\begin{eqnarray}
\label{def:G1G2}
&&\langle N^*(P') | j_{\nu}^{\rm em} |N(P)\rangle \,=\,
 \bar{u}_{N^*}(P')\gamma_5 \Gamma_\nu u_N(P)\,,
\nonumber\\
&& \Gamma_\nu \,=\,\frac{{G_1(q^2)}}{m_N^2}(\slashed{q} q_\nu - q^2 \gamma_\nu)
 -i \frac{{G_2(q^2)}}{m_N} \sigma_{\nu\rho} q^{\rho} \,,
\end{eqnarray}
where $q=P'-P$ is the momentum transfer. In what follows we use the standard notation $Q^2=-q^2$.
The helicity amplitudes
$A_{1/2}(Q^2)$ and $S_{1/2}(Q^2)$ for the electroproduction of $N^*(1535)$
can be expressed in terms of the form factors \cite{Aznauryan:2008us}:
\begin{align}
A_{1/2} &= e\, B\,
\Big[ Q^2 G_1(Q^2) + m_N(m_{N^*}-m_N)G_2(Q^2) \Big],
\nonumber \\
{S}_{1/2}&=  \frac{e BC }{\sqrt{2}}
\Big[(m_{N}\!-\!m_{N^*})G_1(Q^2)+m_{N}G_2(Q^2)\Big].
\label{def:A12S12}
\end{align}
Here $e=\sqrt{4\pi\alpha}$ is the elementary charge and $B$, $C$ are kinematic factors defined as
\begin{eqnarray}
   B &=& \sqrt{\frac{Q^2+(m_{N^*}+m_N)^2}{2 m_N^5(m_{N^*}^2-m_N^2)}} \,,
\nonumber\\
   C & =& \sqrt{1+\frac{(Q^2-m_{N^*}^2+m_N^2)^2}{4 Q^2 m_{N^*}^2}} \,.
\end{eqnarray}

\begin{figure}[ht]
\centerline{\includegraphics[width=4cm, clip = true]{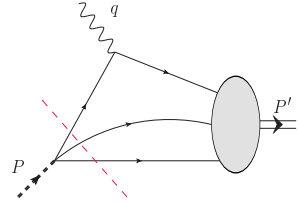}}
\caption{\label{figsum}\small
Schematic structure of the light-cone sum rule for electroproduction form factors.
}
\end{figure}

The basic object of the LCSR approach to baryon form factors~\cite{Braun:2001tj,Braun:2006hz}
is the correlation function
\begin{align}
T_\nu(P,q) &= i\int\! dx\, e^{-iqx}\langle N^\ast(P')| T \{ j_\nu(x) \eta_N(0) \} | 0 \rangle \
\label{CF}
\end{align}
in which
$j$  represents the electromagnetic (or weak) probe and $\eta_N$
is a suitable local operator with nucleon quantum numbers.
 The $N^\ast$ resonance  is explicitly represented by its state vector
 $\langle N^\ast(P') | $, see a schematic representation in Fig.~\ref{figsum}.
The LCSR is obtained by comparing (matching) two different representations for
the correlation function. On the one hand, when both  the momentum transfer  $q^2=-Q^2$ and
 the momentum $P^2 = (P'-q)^2$ flowing in the $\eta_N$ vertex are large and negative, the
 main contribution to the integral comes from the light-cone region $x^2\to 0$ and
 can be studied using the operator product expansion (OPE) of the time-ordered product
$T \{ j(x) \eta_N(0) \}$.  The singularity at $x^2\to 0$ of a particular contribution
is governed by the twist of the relevant composite operator whose matrix element
$\langle N^\ast|\ldots| 0 \rangle $ is related to the $N^\ast$ DA.
On the other hand, one can represent the answer in form of the dispersion integral
in $P^2$ and define the nucleon contribution by the cutoff in the quark-antiquark invariant mass,
the so-called interval of duality $s_0$ (or continuum threshold).
The main role of the interval of duality is that it does not allow large momenta $|k^2| > s_0$ to flow
through the $\eta_N$-vertex; to the lowest order $O(\alpha_s^0)$ one obtains a purely soft
contribution to the form factor as a sum of terms ordered by twist of the relevant operators and
hence including both the leading- and the higher-twist nucleon DAs.
Note that the contribution of higher-twist DAs is suppressed by powers of
the continuum threshold (or by powers of the Borel parameter after applying the
usual QCD sum rule machinery), but not by powers of $Q^2$, the reason being that
soft contributions are not constrained to small transverse separations.

The ``plus'' spinor projection (\ref{projectors}) of the correlation function (\ref{CF})
involving the ``plus'' component of the electromagnetic current
can be parametrized in terms of two invariant functions
\begin{equation}
\label{project4}
  \Lambda_+ T_+ = p_+\left\{ m_N \mathcal{A}(Q^2,P'^2)  +
    \slashed{q}_\perp \mathcal{B}(Q^2,P'^2)\right\}N^+(P)\,,
\end{equation}
where $Q^2=-q^2$ and $P'^2 = (P-q)^2$.
The correlation functions $\mathcal{A}(Q^2,P'^2)$ and $\mathcal{B}(Q^2,P'^2)$
can be calculated in QCD in terms of $N^*$ DAs for sufficiently large Euclidean momenta
$Q^2, -P'^2 \gtrsim 1$~GeV$^2$ using OPE. Schematically,
\begin{eqnarray}
\mathcal{A}(Q^2,P'^2)  &\!=\!& \sum_k \int\! [dx] a_k(Q^2,P'^2\!,x_i,\mu_F^2) F_k(x_i,\mu_F^2),
\nonumber\\
\mathcal{B}(Q^2,P'^2)  &\!=\!& \sum_k \int\! [dx] b_k(Q^2,P'^2\!,x_i,\mu_F^2) F_k(x_i,\mu_F^2),
\end{eqnarray}
where the sum goes over all existing DAs, $F_k \in \{ V_k, A_k, T_k, S_k, P_k\}$ defined
in Eq.~(\ref{defdisampstar}), the integration goes over quark momentum fractions and
$\mu_F$ stands for the factorization scale. The coefficient functions $a_k(Q^2,P'^2,x_i,\mu_F^2)$
and $b_k(Q^2,P'^2,x_i,\mu_F^2)$ are known to the NLO accuracy for twist-three and twist-four DAs~\cite{Anikin:2013aka},
and to leading order (LO) for twist-five and twist-six. In principle this expansion also contains
contributions of four-particle DAs with an additional gluon, five-particle with two gluons or a quark-antiquark pair, etc.
Such contributions start at twist-four and they are not included in the present calculation because the corresponding DAs
are very poorly known (see, however, Ref.~\cite{Braun:2011aw}). It turns out that the coefficient functions are the
same for the states with negative and positive parity, $N^\ast(1535)$ and the nucleon, if the definitions are chosen as
explained in Appendix~A. Thus we are able to use the NLO expressions for the
electromagnetic nucleon form factors obtained in~\cite{Anikin:2013aka} with trivial modifications, e.g.
replacing nucleon mass $m_N$ by $m_{N^\ast}$. One difference is that, because of the larger mass, corrections
of the type $m^2_{N^\ast}/Q^2$ become much larger and numerically significant. For this reason in this work
we use complete expressions for the LO coefficient functions from Ref.~\cite{Braun:2006hz} rather than
the corresponding expressions from Ref.~\cite{Anikin:2013aka} where the expansion in powers of
$m^2_{N^\ast}/Q^2$ was truncated to match the accuracy of the calculated NLO corrections.

The results of the QCD calculation in Euclidean region can be presented in the form of a dispersion relation
\begin{eqnarray}
 \hspace*{-1mm}\mathcal{A}^{\rm QCD}(Q^2,P'^2) &\!=\!& \frac1{\pi}\int_0^\infty\!\!\frac{ds}{s-P'^2} \text{Im}\, \mathcal{A}^{\rm QCD}(Q^2,s)+\ldots
\nonumber\\
 \hspace*{-1mm}\mathcal{B}^{\rm QCD}(Q^2,P'^2) &\!=\!& \frac1{\pi}\int_0^\infty\!\!\frac{ds}{s-P'^2} \text{Im}\, \mathcal{B}^{\rm QCD}(Q^2,s)+\ldots
\end{eqnarray}
where the ellipses indicate possible subtractions.
The same correlation functions can be written in terms of physical spectral densities that
contain a nucleon (proton) pole at $P'^2 \to m_N^2$, nucleon resonances and the continuum.
The nucleon contribution is, obviously, proportional to the electroproduction form factors of interest,
whereas for higher mass states one can use quark-hadron duality:
\begin{eqnarray}
  \hspace*{-1mm}\mathcal{A}^{\rm phys}(Q^2,P'^2) &=& \frac{2\lambda_1 F_1(Q^2)}{m_N^2-P'^2}
\nonumber\\&+&
\frac1{\pi}\int_{s_0}^\infty\!\!\frac{ds}{s-P'^2} \text{Im}\, \mathcal{A}^{\rm QCD}(Q^2,s)+\ldots
\nonumber\\
  \hspace*{-1mm}\mathcal{B}^{\rm phys}(Q^2,P'^2) &=& \frac{\lambda_1 F_2(Q^2)}{m_N^2-P'^2}
\nonumber\\&+&
\frac1{\pi}\int_{s_0}^\infty\!\!\frac{ds}{s-P'^2} \text{Im}\, \mathcal{B}^{\rm QCD}(Q^2,s)+\ldots
\end{eqnarray}
where $s_0\simeq (1.5$~GeV$)^2$ is the interval of duality (also called continuum threshold).
Matching the two above representations and making the Borel transformation that
eliminates subtractions constants
\begin{equation}
  \frac{1}{s-P'^2} \longrightarrow e^{-s/M^2}
\end{equation}
one obtains the sum rules
\begin{align}
\frac{2\lambda_1^N Q^2 G_1(Q^2)}{m_N m_{N^\ast}}
&= \frac1{\pi}\int_0^{s_0}ds \, e^{(m_N^2-s)/M^2} \text{Im}\, \mathcal{A}^{\rm QCD}(Q^2,s)\,,
\nonumber\\
-2 \lambda_1^N G_2(Q^2)
&= \frac1{\pi}\int_0^{s_0}ds \, e^{(m_N^2-s)/M^2} \text{Im}\, \mathcal{B}^{\rm QCD}(Q^2,s)\,.
\label{eq:LCSRscheme}
\end{align}
The dependence on the Borel parameter $M^2$ is unphysical and has to disappear in the full QCD calculation.
It can be used to estimate theoretical uncertainties.

\section{Numerical analysis}


\begin{table*}[t]
\renewcommand{\arraystretch}{1.2}
\begin{center}
\begin{tabular}{@{}l|l|l|l|l|l|l|l|l|l|l@{}} \hline
Method     & $\lambda^{N}_1/\lambda^{N^\ast}_1 $ &$f_{N^\ast}/\lambda^{N^\ast}_1 $ & $\varphi_{10}$ & $\varphi_{11}$ & $\varphi_{20}$ & $\varphi_{21}$ & $\varphi_{22}$ & $\eta_{10}$   & $\eta_{11}$    & Reference \\ \hline
LCSR (1) & 0.633     &0.027      & 0.36     & -0.95      & 0       & 0        & 0         & 0.00    & 0.94   & this work \\ \hline
LCSR (2) & 0.633     &0.027      & 0.37     & -0.96      & 0       & 0        & 0         & -0.29   & 0.23   & this work \\ \hline
LATTICE  & 0.633(43) &0.027(2)   & 0.28(12)  & -0.86(10)  & 1.7(14) & -2.0(18) & 1.7(26)   & -       & -      & \cite{Braun:2014wpa} \\ \hline
\end{tabular}
\end{center}
\caption[]{\small\sf Parameters of the $N^\ast(1535)$ distribution amplitudes at the scale $\mu^2=2$~GeV$^2$.
              For the lattice results \cite{Braun:2014wpa} only statistical errors are shown.
The set of parameters indicated as LCSR~(1) corresponds to the fit to the form factors $G_1(Q^2)$ and $G_2(Q^2)$
extracted from the measurements of helicity amplitudes in Ref.~\cite{Aznauryan:2009mx}
adding the errors in quadrature. The set of parameters indicated as LCSR~(2) is obtained from the fit to helicity amplitudes including
all available data at $Q^2 \ge 1.7~\text{GeV}^2$~\cite{Denizli:2007tq,Dalton:2008aa,Armstrong:1998wg,Aznauryan:2009mx}.
 }
\label{tab:shape}
\renewcommand{\arraystretch}{1.0}
\end{table*}

Main nonperturbative input to the LCSRs for electroproduction form factors
is provided by the DAs of nucleon resonances that can be parameterized by
two normalization constants $f_{N^\ast}$, $\lambda_1^{N^\ast}$
and a set of shape parameters $\varphi_{nk}$, $\eta_{nk}$ corresponding to
contributions of local operators of increasing dimension, see Eqs.~(\ref{expand-varphi}), (\ref{PhiPsi4}).
The dependence of the form factors on these parameters is linear so that the results can
conveniently be presented as
\begin{align}
 G_1(Q^2) &= \frac{\lambda_1^{N^\ast}}{\lambda_1^N} \biggl\{ g^{00}_{1}(Q^2) + g^{10}_{1}(Q^2)\eta_{10} +  g^{11}_{1}(Q^2)\eta_{11}
\nonumber\\&\hspace*{-0.8cm}{}+
  \frac{f_{N^\ast}}{\lambda_1^{N^\ast}}\Big[ f_1^{00}(Q^2) + f_1^{10}(Q^2)\varphi_{10}  + f_1^{11}(Q^2)\varphi_{11}+\ldots\Big]\biggr\}
\label{G1param}
\end{align}
and similarly
\begin{align}
 G_2(Q^2) &= \frac{\lambda_1^{N^\ast}}{\lambda_1^N} \biggl\{ g^{00}_{2}(Q^2) + g^{10}_{2}(Q^2)\eta_{10} +  g^{11}_{2}(Q^2)\eta_{11}
\nonumber\\&\hspace*{-0.8cm}{}+
  \frac{f_{N^\ast}}{\lambda_1^{N^\ast}}\Big[ f_2^{00}(Q^2) + f_2^{10}(Q^2)\varphi_{10}  + f_2^{11}(Q^2)\varphi_{11}+\ldots\Big]\biggr\}
\label{G2param}
\end{align}
where the ellipses stand for the contributions of second-order polynomials in the leading-twist DAs (\ref{expand-varphi}), terms in
$\varphi_{20}$, $\varphi_{21}$, $\varphi_{22}$.
The coefficient functions $f_{1,2}^{n,k}(Q^2)$ and  $g_{1,2}^{n,k}(Q^2)$ are given by very cumbersome analytic expressions~\cite{Anikin:2013aka}
and depend implicitly on the masses of $N^\ast$ and the nucleon, the continuum threshold $s_0$, Borel parameter $M^2$, QCD
coupling $\alpha_s(\mu_F)$ and the factorization scale $\mu_F$. Note that, e.g., $f_1^{00}(Q^2)$ includes the sum of contributions
of the asymptotic leading-twist DA and the corresponding Wandzura-Wilczek terms in the higher-twist DAs,
see Appendix A and Ref.~\cite{Anikin:2013aka} for more details. Note also that the DA $\Xi_4$ (\ref{Xi4star}) corresponding to the
$\ell_z=2$ component of the light-front three-quark wave function does not contribute to the LCSRs for our choice of the
nucleon interpolating current.

Calculations in this work are done for the "standard" choice of the specific LCSR parameters:
continuum threshold $s_0= (1.5~\text{GeV})^2$, Borel parameter $M^2 = 2~\text{GeV}^2$ and factorization (and renormalization)
scale $\mu_F^2 = 2~\text{GeV}^2$. The dependence on these parameters is rather mild; in particular varying the Borel parameter in
the range $1.5-2$~GeV$^2$ induces an overall variation of form factor of the order of 10\% so that, e.g., the ratio $G_2/G_1$ is
largely unchanged.

The resonance mass corrections enter the LCSRs in a complicated way, as terms in $m_{N^\ast}^2/Q^2$ and $m_{N^\ast}^2/s_0$.
The latter ones do not decrease at large momentum transfers and in ideal case have to be resummed to all orders.
The corresponding expression exists for the LO LCSRs~\cite{Braun:2006hz} but not for the NLO corrections.
In order to minimize this mismatch we have rescaled the $\mathcal{O}(\alpha_s)$ contributions calculated in~\cite{Anikin:2013aka}
by the ratio of the corresponding LO terms calculated with account for $m^2_{N^\ast}$ corrections and putting $m^2_{N^\ast}$ to zero.
For the numerically important contributions this rescaling corresponds to a reduction of the NLO correction by $10-20\%$.

Existing information on the DAs of negative parity resonances is very scarce.
The results of the recent lattice calculation~\cite{Braun:2014wpa} are presented in Table~1. The most interesting feature
of these results is that the corrections to the asymptotic leading twist DAs have alternating signs
for the lattice states with increasing mass. In particular the twist-three DA of $N^*(1535)$ has a very small value at the origin
and is approximately antisymmetric with respect to the exchange of the two valence quarks forming a scalar ``diquark'', whereas
the DA of $N^*(1535)$ is symmetric and similar in shape to the nucleon DA, see Fig.~9 in Ref.~\cite{Braun:2014wpa}.
These results are still exploratory and have to be taken with caution because identification of lattice states with particular
physical resonances is not obvious and requires further study.
Even with this uncertainty, the lattice values are very helpful as knowing the order of magnitude of the parameters
allows one to establish a hierarchy of different contributions to the LCSR.

As an illustration, the NLO LCSR result for the form factors at $Q^2=2$~GeV$^2$ normalized to the dipole formula
\begin{align}
 D(Q^2) &= \frac{1}{(1+Q^2/a)^2}\,,\qquad a = 0.71~\text{GeV}^2
\label{dipole}
\end{align}
can be written as follows:
\begin{align}
 \frac{G^{\rm NLO}_1(Q^2)}{D(Q^2)} =& \frac{\lambda_1^{N^\ast}}{\lambda_1^N}
\Big[ 0.666
-2.18 \eta_{10}
+0.86 \eta_{11}
\nonumber\\[-1mm]&{}
-0.69 \tilde f_{N^\ast}
-1.76 \tilde f_{N^\ast} \varphi_{10}
+1.05 \tilde f_{N^\ast} \varphi_{11}
\nonumber\\&{}
 +1.3 \tilde f_{N^\ast} \varphi_{20}
 +0.66 \tilde f_{N^\ast} \varphi_{21}
- 0.06 \tilde f_{N^\ast} \varphi_{22}
\Big],
\nonumber\\
 \frac{G^{\rm NLO}_2(Q^2)}{D(Q^2)} =& \frac{\lambda_1^{N^\ast}}{\lambda_1^N}
\Big[ - 0.466
+ 1.84 \eta_{10}
+0.06 \eta_{11}
\nonumber\\[-1mm]&{}
-0.82 \tilde f_{N^\ast}
-1.06 \tilde f_{N^\ast} \varphi_{10}
-1.08 \tilde f_{N^\ast} \varphi_{11}
\nonumber\\&{}
 +2.6 \tilde f_{N^\ast} \varphi_{20}
 +1.5 \tilde f_{N^\ast} \varphi_{21}
 + 0.39 \tilde f_{N^\ast} \varphi_{22}
\Big]
\nonumber
\end{align}
where we use a notation $\tilde f_{N^\ast} $ for the ratio of twist-three and twist-four couplings
\begin{align}
   \tilde f_{N^\ast} = \frac{f_{N^\ast}}{\lambda_1^{N^\ast}}  = 0.027(2)\quad \text{\cite{Braun:2014wpa}}\,.
\label{tildefN}
\end{align}
For comparison, the similar decomposition of the form factors for the LO LCSRs~\cite{Braun:2009jy}
for the same value $Q^2=2$~GeV$^2$ reads
\begin{align}
 \frac{G^{\rm LO}_1(Q^2)}{D(Q^2)} =& \frac{\lambda_1^{N^\ast}}{\lambda_1^N}
\Big[ 0.816
-2.02 \eta_{10}
+0.88 \eta_{11}
\nonumber\\[-1mm]&{}
-0.59 \tilde f_{N^\ast}
-1.60 \tilde f_{N^\ast} \varphi_{10}
+1.19 \tilde f_{N^\ast} \varphi_{11}
\nonumber\\&{}
 +1.26 \tilde f_{N^\ast} \varphi_{20}
 +0.70 \tilde f_{N^\ast} \varphi_{21}
 +0.12 \tilde f_{N^\ast} \varphi_{22}
\Big],
\nonumber\\
 \frac{G^{\rm LO}_2(Q^2)}{D(Q^2)} =& \frac{\lambda_1^{N^\ast}}{\lambda_1^N}
\Big[ - 0.466
+ 1.84 \eta_{10}
+0.06 \eta_{11}
\nonumber\\[-1mm]&{}
-1.19 \tilde f_{N^\ast}
-0.78 \tilde f_{N^\ast} \varphi_{10}
+3.82 \tilde f_{N^\ast} \varphi_{11}
\nonumber\\&{}
 +2.9 \tilde f_{N^\ast} \varphi_{20}
 +1.6 \tilde f_{N^\ast} \varphi_{21}
 + 0.28 \tilde f_{N^\ast} \varphi_{22}
\Big]
\nonumber
\end{align}
so that the NLO corrections are significant.

\begin{figure*}[t]
   \begin{center}
\epsfig{file=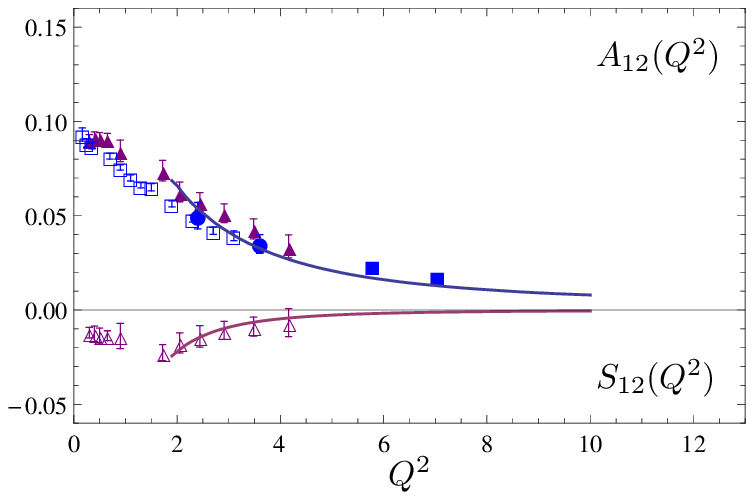,scale=1}
\qquad
\epsfig{file=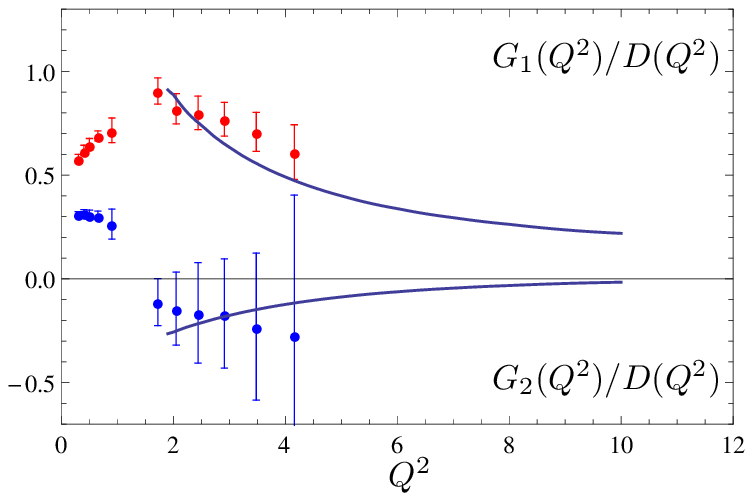,scale=1}
   \end{center}
\caption{\small\sf Helicity amplitudes $A_{12}$ and $S_{12}$ for electroproduction of $N^*(1535)$ (left panel) and the form factors $G_1(Q^2)$,
$G_2(Q^2)$, normalized to the dipole formula (right panel).
Experimental data on the left panel are taken from \cite{Denizli:2007tq} (empty squares) \cite{Dalton:2008aa} (filled squares)
\cite{Armstrong:1998wg} (filled circles) and \cite{Aznauryan:2009mx} (triangles).
The form factors on the right panel are calculated from the data~\cite{Aznauryan:2009mx} on helicity amplitudes
adding the errors in quadrature.
The curves show the results of the NLO LCSR fit to the form factors $G_1(Q^2)$ and $G_2(Q^2)$ for $Q^2 \ge 1.7~\text{GeV}^2$
with parameters of the $N^\ast(1535)$ DAs specified in the first line in Table 1.
}
\label{fig:ABO1}
\end{figure*}
\begin{figure*}[ht]
   \begin{center}
\epsfig{file=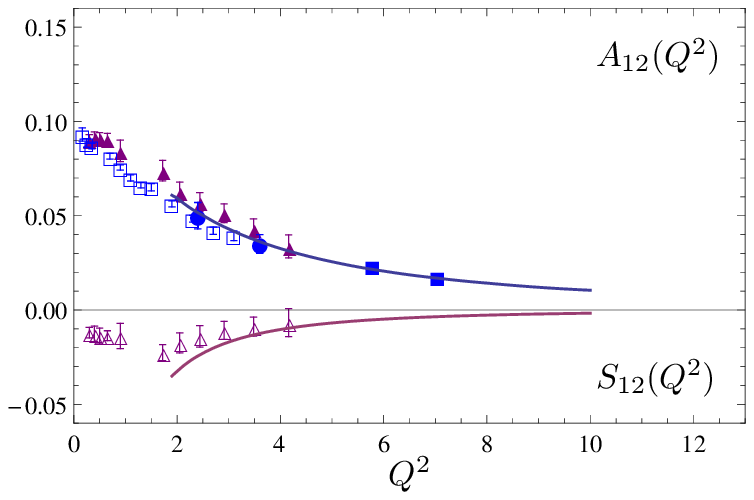,scale=1}
\qquad
\epsfig{file=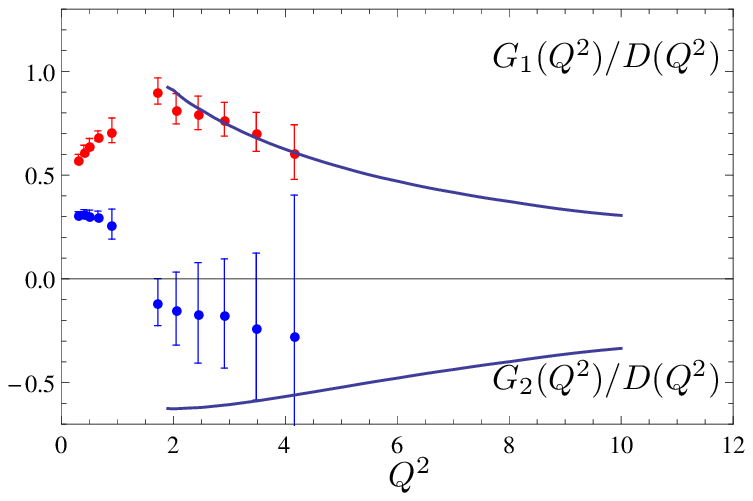,scale=1}
   \end{center}
\caption{\small\sf The same as in Fig.~\ref{fig:ABO1} but for the fit to helicity amplitudes $A_{12}$, $S_{12}$ including all available data at $Q^2 \ge 1.7~\text{GeV}^2$.
The fitted parameters of the $N^\ast(1535)$ DAs are specified in the second line in Table~1.
}
\label{fig:ABO2}
\end{figure*}
For convenience we provide a simple parametrization for the coefficient
functions $f^{nk}_{1,2}$, $g^{nk}_{1,2}$ (\ref{G1param}), (\ref{G2param})
as functions of $Q^2$ in Appendix~B.
This parametrization was obtained for the region of momentum transfers $2~\text{GeV}^2 < Q^2 < 12~\text{GeV}^2$
and should not be used outside this interval. In particular we found that the
mass corrections $\sim m_{N^\ast}^2/Q^2$ become very large for $Q^2 < 2~\text{GeV}^2$ so that
the LCSRs become unstable (and not reliable).
In general, different contributions to the LCSRs are distinguished by their $Q^2$-dependence so that
one needs a sufficient lever arm in $Q^2$ to determine several of them simultaneously.

Since the existing data for $Q^2\ge  1.5-2~\text{GeV}^2$ are very limited, 
we put in this work all second-order coefficients in the leading-twist DAs to zero,
$\varphi_{20}=\varphi_{21}=\varphi_{22} =0$, used central lattice values for $f_{N^\ast}$ and $\lambda_1^{N^\ast}$, 
and constrained $\varphi_{10}, \varphi_{11}$ to the lattice values within the given error bars. 
In this way we are left, essentially, with two free parameters --- $\eta_{10}$ and $\eta_{11}$.    
We expect that much more data will become available after the 12 GeV upgrade at  
Jefferson Lab where a dedicated experiment is planned to study electroproduction of nucleon resonances
at high photon virtualities up to $Q^2 = 12$~GeV$^2$~\cite{Aznauryan:2012ba}.


Information on the electrocouplings of nucleon resonances at large momentum transfers is obtained
by studying electroproduction of $\pi$ and $\eta$ mesons in the respective resonance
region~\cite{Denizli:2007tq,Dalton:2008aa,Armstrong:1998wg,Aznauryan:2009mx}.
The results are usually presented for the helicity amplitudes, and in earlier work only the larger one, $A_{12}(Q^2)$,
was studied for large momentum transfers.
The latest study~\cite{Aznauryan:2009mx} also includes the results on $S_{12}(Q^2)$ up to $Q^2 = 4.16$~GeV$^2$
allowing us to extract from these data the Dirac-like and Pauli-like transition form factors $G_1(Q^2)$
and $G_2(Q^2)$ (\ref{def:G1G2}) that are more relevant for QCD studies. In this extraction we assumed that
the errors for helicity amplitudes given in Ref.~\cite{Aznauryan:2009mx} are uncorrelated and added them in quadrature.
The results are shown in Fig.~\ref{fig:ABO1} and  Fig.~\ref{fig:ABO2} on the right panels;  it is seen that
the Pauli-like form factor changes sign and becomes negative at large $Q^2$, although the errors are quite large.

Two different LCSR fits of the experimental data are shown in Figs.~\ref{fig:ABO1},~\ref{fig:ABO2}.
The difference is that in Fig.~\ref{fig:ABO1} the fit is done to the form factors extracted from the data on helicity amplitudes
reported in Ref.~\cite{Aznauryan:2009mx}, and in  Fig.~\ref{fig:ABO2} we make a fit to the data on helicity amplitudes $A_{12}(Q^2)$
and $S_{12}(Q^2)$ themselves including all existing data for $Q^2\ge 1.7$~GeV$^2$. In the second case the fit is driven by the data
\cite{Denizli:2007tq,Dalton:2008aa,Armstrong:1998wg} on  $A_{12}(Q^2)$ that have smaller errors and not entirely consistent
with~\cite{Aznauryan:2009mx}, so that a worse description of the form factors in this fit is not a surprise.
The corresponding parameters are listed in Table~1.

Because of the small value of the leading twist normalization constant suggested by lattice calculations (\ref{tildefN}), the results
for $A_{12}(Q^2)$ and $G_1(Q^2)$ prove to be almost insensitive to the leading twist DA of the $N^*(1535)$ resonance
and are dominated by the twist-four contributions corresponding to the P-wave parts of the three-quark light-front wave functions
(see Appendix A). Moreover, sensitivity of the results to the shape parameters of the twist-four DAs, $\eta_{10}$ and $\eta_{11}$,
is rather mild, cf. two last columns in the first and the second  line in Table~1. Thus  $A_{12}(Q^2)$ and $G_1(Q^2)$  are both
sensitive mostly to the ratio of the normalization constants $\lambda^{N}_1/\lambda^{N^\ast}_1$
which we fix to the lattice value $0.633$~\cite{Braun:2014wpa}. The LCSR predictions for these observables are very stable,
 and the agreement of the existing data with the normalization suggested by lattice calculations is encouraging.

The $S_{12}(Q^2)$ amplitude and especially the Pauli-like form factor $G_2(Q^2)$ are much more sensitive to the
nonperturbative input and in particular
to the shape parameters of the twist-four DAs, compare Fig.~\ref{fig:ABO1} and Fig.~\ref{fig:ABO2}.
Also the leading-twist contributions play some role in this case because of strong cancellations.
More precise data and a larger interval in $Q^2$ are needed to make this comparison quantitative.

\section{Conclusions and Outlook}
In this work we argue that the LCSR approach can provide one with quantitative
information on the wave functions of nucleon resonances at short distances.
The basic idea behind this technique is that soft Feynman contributions to the form factors
are calculated in terms of small transverse distance quantities using dispersion relations and duality.
The form factors are thus expressed in terms of light-front wave functions at small
transverse separations, called DAs, without additional parameters.
Alternatively, the distribution amplitudes can be extracted from the comparison with the
experimental data on form factors and compared to the results of lattice QCD simulations or other nonperturbative
approaches based on, e.g.,  QCD sum rules or Dyson-Schwinger equations.
The results of the corresponding NLO calculation for the particular case of the
$N^\ast(1535)$ resonance are presented and compared with the
existing data. We find that the form factors are dominated by twist-four DAs
that are related to the $P$-wave three-quark wave functions, i.e., to the distribution of orbital
angular momentum.

Interestingly enough the LCSRs have the same form for spin-1/2 resonances of both parities so that
apart from the (calculable) effects of resonance mass corrections
the difference in observed form factors of, say, $N^\ast(1535)$ and $N^\ast(1650)$ can be attributed to the
difference in the wave functions, which is of major interest. The differences between nucleon elastic form factors
and electroexcitation of the Roper resonance can be studied in a similar manner; however, it is likely that in the
latter case interpretation of the results may require a better understanding and more sophisticated models
of twist-five DAs than are available at present.

\section*{Acknowledgments}
V.B. is grateful to A.~Belitsky for the correspondence concerning the
relation of light-front wave functions and DAs.
The work by I.A. was partially supported by the Heisenberg-Landau Program of the
German Research Foundation (DFG).


\appendix

\renewcommand{\theequation}{\Alph{section}.\arabic{equation}}

\section*{Appendices}

\section{Light-front wave functions and DAs of $J^P = \left(\frac{1}{2}\right)^-$ nucleon resonances }

In the light-front description~\cite{Lepage:1979zb} a hadron is
represented by the superposition of Fock states with  different number of partons.
Restricting ourselves to the three-quark (valence) components we view, e.g.,
the proton with positive helicity as a superposition of states with different values
of the quark orbital angular momentum projection on the direction of motion, $\ell_z = -1,0,1,2$,
\begin{align}
 |N\up\rangle &= \sum_{\ell_z} |N\up\rangle_{uud}^{\ell_z}\,.
\end{align}
A nonzero value of $\ell_z$ accounts for the mismatch between the proton helicity
and the sum of helicities of the valence quarks $\lambda_i$
so that $1/2 = \lambda_1 +\lambda_2 + \lambda_3 +\ell_z $.
The four different contributions can be written in terms of six independent
scalar light-front wave functions as~\cite{Ji:2002xn,Ji:2003yj,Belitsky:2005qn}
\nopagebreak
\begin{widetext}
\begin{align}\label{NucleonLCWF}
|N\up\rangle_{uud}^{\ell_z=0}&=\frac{\epsilon^{ijk}}{\sqrt{6}}\int\frac{[dx][dk_\perp]}{\sqrt{x_1 x_2 x_3}}
\Big[\psi_{N;1}^{(0)}(1,2,3) + i\epsilon^{\alpha\beta} k^\perp_{1\alpha}{k}^\perp_{2\beta}\psi_{N;2}^{(0)}(1,2,3)\Big]
b_{u\uparrow}^{\dagger i}(1) \Bigl( b_{u\downarrow}^{\dagger j}(2)b_{d\uparrow}^{\dagger k}(3)- b_{d\downarrow}^{\dagger j}(2)b_{u\uparrow}^{\dagger k}(3)
\Bigr)|0\rangle\,,
\notag\\
|N\up\rangle_{uud}^{\ell_z=1}&= \frac{\epsilon^{ijk}}{\sqrt{6}}\int \frac{[dx][dk_\perp]}{\sqrt{x_1 x_2 x_3}}
\Big[k^\perp_1\psi_{N;1}^{(1)}(1,2,3) + k^\perp_2 \psi_{N;2}^{(1)}(1,2,3)\Big]
\Bigl(b_{u\uparrow}^{\dagger i}(1) b_{u\downarrow}^{\dagger j}(2)b_{d\downarrow}^{\dagger k}(3)-
b_{d\uparrow}^{\dagger i}(1)b_{u\downarrow}^{\dagger j}(2)b_{u\downarrow}^{\dagger k}(3)
\Bigr)|0\rangle\,,
\notag\\
|N\up\rangle_{uud}^{\ell_z=-1}&=
\frac{\epsilon^{ijk}}{\sqrt{6}}\int\frac{[dx][dk_\perp]}{\sqrt{x_1 x_2 x_3}}\Big[\bar k^\perp_1 \psi_{N}^{(-1)}(1,2,3)\Big]
b_{u\uparrow}^{\dagger i}(1) \Bigl( b_{u\uparrow}^{\dagger j}(2)b_{d\uparrow}^{\dagger k}(3)-b_{d\uparrow}^{\dagger j}(2)b_{u\uparrow}^{\dagger k}(3)\Bigr)|0\rangle\,,
\notag\\
|N\up\rangle_{uud}^{\ell_z=2}&=
\frac{\epsilon^{ijk}}{\sqrt{6}}\int\frac{[dx][dk_\perp]}{\sqrt{x_1 x_2 x_3}}
\Big[k^\perp_1 k^\perp_3  \psi_{N}^{(2)}(1,2,3)\Big]
\Bigl(b_{u\downarrow}^{\dagger i}(1) b_{u\downarrow}^{\dagger j}(2)b_{d\downarrow}^{\dagger k}(3)-
b_{u\downarrow}^{\dagger i}(1)b_{d\downarrow}^{\dagger j}(2)b_{u\downarrow}^{\dagger k}(3)\Bigr)|0\rangle\,.
\end{align}
\end{widetext}
Here $b_{u\uparrow}^{\dagger i}(1)$ etc. are creation operators for the quarks of
specific flavor with positive $\up$ or negative $\down$ helicity; the argument
$(1)$ stands for the dependence on longitudinal momentum fractions
and transverse momenta of the given quark, i.e. $u_{\uparrow i}(1)=u_{\uparrow i}(x_1,k^\perp_{1})$, and so on.
We use the notation for transverse momenta
\begin{align}
 k^\perp = k^\perp_x + ik^\perp_y\,, && \bar k^\perp = k^\perp_x - i k^\perp_y \,.
\end{align}
The light-front wave functions $\psi_{N;i}^{(\ell_z)}(1,2,3)$ depend on momentum fractions
$x_i$ and transverse momenta squared $|k_{\perp,i}|^2 = k_i\bar k_i$ of all partons.
The integration measure is chosen as~\cite{Braun:2011aw}
\begin{align}\label{measure1}
{}[dx]=&\prod_{k=1}^3 dx_k\delta(1-\sum x_k)\,,
\end{align}
and
\begin{align}\label{measure2}
{}[dk_\perp]=&\frac{1}{4(2\pi)^{6}}\prod_{k=1}^3 d^2k^\perp_{k}\delta^{(2)}\left(\sum k^\perp_{i}\right)\,.
\end{align}
The proton light-cone DAs, in turn, are defined as matrix elements of gauge-invariant nonlocal
operators with the three quark fields separated by a light-like distance.
Standard decomposition \cite{Braun:2000kw} involves 24 invariant functions:
\begin{widetext}
\begin{eqnarray}
\label{defdisamp}
\lefteqn{ 4 \langle 0| \epsilon^{ijk} u_\alpha^i(a_1 n) u_\beta^j(a_2 n) d_\gamma^k(a_3 n)
|N(P,\lambda)\rangle =}
\nonumber \\
&=&
S^N_1 m_N C_{\alpha \beta} \left(\gamma_5 u_N^+\right)_\gamma +
S^N_2 m_N C_{\alpha \beta} \left(\gamma_5 u_N^-\right)_\gamma +
P^N_1 m_N \left(\gamma_5 C\right)_{\alpha \beta} (u^+_N)_\gamma +
P^N_2 m_N \left(\gamma_5 C \right)_{\alpha \beta} (u_N^-)_\gamma
\nonumber \\
&& +
V^N_1  \left(\!\not\!{p}C \right)_{\alpha \beta} \left(\gamma_5 u_N^+\right)_\gamma +
V^N_2  \left(\!\not\!{p}C \right)_{\alpha \beta} \left(\gamma_5 u_N^-\right)_\gamma +
\frac12 V^N_3 m_N  \left(\gamma_\perp C \right)_{\alpha \beta}\left( \gamma^{\perp}
\gamma_5 u_N^+\right)_\gamma
\nonumber \\
&& +
\frac12 V^N_4 m_N  \left(\gamma_\perp C \right)_{\alpha \beta}\left( \gamma^{\perp}
\gamma_5 u_N^-\right)_\gamma +
V^N_5  \frac{m_N^2}{2 p n}
\left(\!\not\!{n}C \right)_{\alpha \beta} \left(\gamma_5 u_N^+\right)_\gamma +
\frac{m_N^2}{2 pn} V^N_6  \left(\!\not\!{n}C \right)_{\alpha \beta} \left(\gamma_5 u_N^-\right)_\gamma
\nonumber \\
&& +
A^N_1  \left(\!\not\!{p}\gamma_5 C \right)_{\alpha \beta} (u_N^+)_\gamma +
A^N_2  \left(\!\not\!{p}\gamma_5 C \right)_{\alpha \beta} (u_N^-)_\gamma +
\frac12 A^N_3 m_N  \left(\gamma_\perp \gamma_5 C \right)_{\alpha \beta}\left( \gamma^{\perp}
u_N^+\right)_\gamma
\nonumber \\
&& +
\frac12 A^N_4 m_N  \left(\gamma_\perp \gamma_5 C \right)_{\alpha \beta}\left( \gamma^{\perp}
u_N^-\right)_\gamma +
A^N_5  \frac{m_N^2}{2 p n}
\left(\!\not\!{n}\gamma_5 C \right)_{\alpha \beta} (u_N^+)_\gamma +
\frac{m_N^2}{2 pn}  A^N_6  \left(\!\not\!{n}\gamma_5 C \right)_{\alpha \beta} (u_N^-)_\gamma
\nonumber \\
&& +
T^N_1 \left(i \sigma_{\perp p} C\right)_{\alpha \beta}
\left(\gamma^\perp\gamma_5 u_N^+\right)_\gamma +
T^N_2 \left(i \sigma_{\perp\, p} C\right)_{\alpha \beta}
\left(\gamma^\perp\gamma_5 u_N^-\right)_\gamma +
T^N_3 \frac{m_N}{p n}
\left(i \sigma_{p\, n} C\right)_{\alpha \beta}
\left(\gamma_5 u_N^+\right)_\gamma
\nonumber \\
&&
+ T^N_4 \frac{m_N}{p n}\left(i \sigma_{n\, p} C\right)_{\alpha \beta}
\left(\gamma_5 u_N^-\right)_\gamma +
T^N_5 \frac{m_N^2}{2 p n}  \left(i \sigma_{\perp\, n} C\right)_{\alpha \beta}
\left(\gamma^\perp\gamma_5 u_N^+\right)_\gamma +
\frac{m_N^2}{2 pn}  T^N_6 \left(i \sigma_{\perp\, n} C\right)_{\alpha \beta}
\left(\gamma^\perp\gamma_5 u_N^-\right)_\gamma
\nonumber \\
&& + \frac12 m_N T^N_7 \left(\sigma_{\perp\, \perp'} C\right)_{\alpha \beta}
\big(\sigma^{\perp\, \perp'} \gamma_5 u_N^+\big)_\gamma +
\frac12 m_N T^N_8  \left(\sigma_{\perp\, \perp'} C\right)_{\alpha \beta}
\big(\sigma^{\perp\, \perp'} \gamma_5 u_N^-\big)_\gamma\,,
\end{eqnarray}
\end{widetext}
In this expression $\alpha,\beta,\gamma$ are spinor indices,
$n_\mu$ is an auxiliary light-like vector, $n^2=0$,
\begin{align}
 p_\mu = P_\mu - \frac12 \frac{m_N^2}{Pn}\,, && p^2 =0\,,
\end{align}
where $P_\mu$ is the proton momentum, $P^2=m_N^2$.
Further, $u_N^\pm = \Lambda_\pm u_N(P,\lambda)$ where $u_N(P,\lambda)$ is the usual Dirac spinor in
relativistic normalization, the
projectors are defined as
\begin{align}
 &\Lambda_+ = \frac{\slashed{p}\slashed{n}}{2pn}\,,
\qquad
 \Lambda_- = \frac{\slashed{n}\slashed{p}}{2pn}\,,
\notag\\
 &  g^\perp_{\mu\nu} = g_{\mu\nu} - \frac{p_\mu n_\nu+p_\nu n_\mu}{pn}
\label{projectors}
\end{align}
and $C$ is the charge-con\-ju\-ga\-tion matrix.
We use a shorthand notation $\sigma_{\perp n} \otimes \gamma^\perp = \sigma_{\mu\nu} n^\nu g^{\mu\alpha}_\perp\otimes \gamma_\alpha$,
etc.
The invariant functions $F = V_i, A_i, T_i$ correspond to contributions of a given
\emph{collinear twist} and can be written as Fourier integrals
\begin{align}
 F(a_j,pn) &= \int [dx]\, e^{-i(pn)\sum_i x_ia_i} F(x_i)
\end{align}
where $F(x_i)$ depend on the three valence quark momentum fractions $x_i$.

Using various symmetries these functions can be combined in 8 independent
light-cone DAs \cite{Braun:2000kw}.
There exists a single DA for the leading twist-three~\cite{Lepage:1979zb}
\begin{align}
\label{NDA:twist3}
\ide{2em} \langle 0 |
\epsilon^{ijk}\! \left(u^{\up}_i(a_1 n) C \!\!\not\!{n} u^{\down}_j(a_2 n)\right)
\slashed{n} d^{\up}_k(a_3 n)
|N(P,\lambda)\rangle \nonumber\\
&= - \frac12 f_N\,(pn) \slashed{n}\, u_N^\up(P)\! \!\int\! [dx]
\,e^{-i (pn) \sum x_i a_i}\,
\varphi_N(x_i)\,,
\end{align}
such that~\cite{Chernyak:1983ej}
\begin{eqnarray}
V_1(1,2,3) &=& \frac12f_N\big[\varphi_N(1,2,3)+\varphi_N(2,1,3)\big]\,,
\nonumber\\
A_1(1,2,3) &=& \frac12f_N\big[\varphi_N(2,1,3)-\varphi_N(1,2,3)\big]\,,
\nonumber\\
T_1(1,2,3) &=& \frac12f_N\big[\varphi_N(1,3,2)+\varphi_N(2,3,1)\big]\,,
\label{VAT1}
\end{eqnarray}
and for twist-four there are three independent DAs~\cite{Braun:2000kw}
\begin{align} \label{NDA:twist4}
\ide{2em} \langle 0 | \epsilon^{ijk}\! \left(u^{\up}_i(a_1 n) C\slashed{n} u^{\down}_j(a_2 n)\right)
\!\slashed{p} d^{\up}_k(a_3 n) |N(P,\lambda)\rangle
\nonumber \\
=& -\frac14(pn)\slashed{p}\, u_{N}^{\up}(P)\! \!\int\! [dx] \,e^{-i (pn) \sum x_i a_i}\,
\nonumber \\
&\times \left[f_{N}\Phi^{N,WW}_4(x_i)+\lambda^N_1\Phi^{N}_4(x_i)\right],
\end{align}
\begin{align}
\ide{2em} \langle 0 | \epsilon^{ijk}\! \left(u^{\up}_i(a_1 n) C \slashed{n}\gamma_{\perp}\slashed{p}
u^{\down}_j(a_2 n)\right) \gamma^{\perp}\slashed{n} d^{\up}_k(a_3 n) |N(P,\lambda)\rangle
\nonumber\\
=& -\frac12 \, (pn) \slashed{n}\,m_{N} u_{N}^{\up}(P)\! \!\int\! [dx] \,e^{-i (p n) \sum x_i a_i}\,
\nonumber \\
&\times \left[f_{N}\Psi^{N,WW}_4(x_i)-\lambda^N_1\Psi^{N}_4(x_i)\right],
\end{align}
\begin{align}
\ide{2em} \langle 0 | \epsilon^{ijk}\! \left(u^{\up}_i(a_1 n) C\slashed{p}\,\slashed{n} u^{\up}_j(a_2 n)\right)
 \slashed{n} d^{\up}_k(a_3 n) |N(P,\lambda)\rangle
\nonumber \\
=& \frac{\lambda^N_2}{12} (pn) \slashed{n}  m_{N} u_{N}^{\up}(P)\! \!\int\! [dx] \,e^{-i (pn) \sum x_i a_i}\,\Xi^{N}_4(x_i)\,,
\end{align}
where $\Phi^{N,WW}_4(x_i)$ and $\Psi^{N,WW}_4(x_i)$ are the so-called
Wandzura-Wilczek contributions that can be expressed in terms of the
leading-twist DA $\varphi_N(x_i)$~\cite{Braun:2008ia,Anikin:2013yoa}.
The constants $f_N$, $\lambda_1^N$ and $\lambda_2^N$
are defined in such a way that the integrals of the
DAs  $\varphi_N$, $\Phi_4$, $\Psi_4$, $\Xi_4$ are normalized to unity:
\begin{align}
 \int [dx]\, F(x_i) = 1\,,\qquad F \in \{\varphi_N,\Phi_4,\Psi_4,\Xi_4\}.
\label{norm}
\end{align}

Using the canonical expansion of the quark fields in (\ref{NDA:twist3}),(\ref{NDA:twist4}) in terms of creation and
annihilation operators and Dirac equation to eliminate ``bad'' quark field components it is easy to calculate the
required matrix elements from the set of light-front
wave functions in Eqs.~(\ref{NucleonLCWF}). In this way one obtains for our normalization
(cf. \cite{Belitsky:2002kj,Belitsky:2005qn})
\begin{widetext}
\begin{eqnarray}
 f_N \varphi_N(x_1,x_2,x_3) &=& -4\sqrt{6} \int [dk_\perp] \,\psi_{N;1}^{(0)}(1,2,3)\,,
\nonumber\\
{}[\lambda^N_1\Phi^{N}_4+ f_{N}\Phi^{N,WW}_4](x_2,x_1,x_3)
&=&
-8\sqrt{6}\int \frac{[dk_\perp]}{x_3 m_N} k^\perp_3\cdot\Big[\bar k^\perp_1\psi_{N;1}^{(1)} + \bar k^\perp_2 \psi_{N;2}^{(1)}\Big](1,2,3)\,,
\nonumber\\
{}[\lambda^N_1\Psi^{N}_4 - f_{N}\Psi^{N,WW}_4](x_1,x_2,x_3)
&=&
-8\sqrt{6} \int \frac{[dk_\perp]}{x_2 m_N} \bar k^\perp_2\cdot\Big[k^\perp_1\psi_{N;1}^{(1)} + k^\perp_2 \psi_{N;2}^{(1)}\Big](1,2,3)\,,
\nonumber\\
\lambda^N_2 \Xi^{N}_4(x_1,x_2,x_3)
&=&
-24\sqrt{6} \int \frac{[dk_\perp]}{x_1 m_N} k^\perp_1\cdot \Big[\bar k^\perp_1\Big(\psi_{N}^{(-1)}(1,3,2)-\psi_{N}^{(-1)}(1,2,3)\Big)
\nonumber\\&&{}\hspace*{2.99cm}
+ \bar k^\perp_2\Big(\psi_{N}^{(-1)}(2,3,1)-\psi_{N}^{(-1)}(2,1,3)\Big)\Big]
\end{eqnarray}
\end{widetext}
so that DAs correspond to integrals over the light-front wave functions over transverse momenta,
with some prefactors. One has to have in mind that these relations are somewhat schematic since
transverse momentum integrals on the right-hand side (r.h.s.) are divergent and have to be regulated e.g.
introducing a cutoff. In turn, the DAs are usually defined using dimensional regularization and the MS-subtraction
so that a matching coefficient can be necessary.
Also the wave function renormalization factors have to be added for the quark fields.
The twist-four DAs include additional contributions from the four-particle Fock states
with an extra gluon~\cite{Ji:2003yj,Braun:2011aw}. If these contributions are taken into account,
the four-particle quark-gluon nucleon DAs have to be added as well~\cite{Braun:2008ia,Braun:2011aw}.

The complete set of nucleon DAs carries the full information on the nucleon structure,
in the same manner as the complete basis of light-front wave functions.
In practice, however, both expansions have to be truncated and the usefulness of a truncated version, taking into account
either the first few Fock states or a few lowest twist
contributions, may depend on the concrete physics application.

The classification of the three-quark nucleon light-front wave functions in Eq.~(\ref{NucleonLCWF}) can be overtaken for the
negative parity isospin-1/2 resonances, e.g. $N^*(1535)$,  without modification. The symmetry under parity transformation
does not constrain the light-front wave functions but affects the relation between the wave functions of the states with opposite helicity
in terms of the helicity-flipped quarks. The corresponding expressions can be worked out using the Jacobi-Wick transformation~\cite{Jacob:1959at}
\begin{align}
    \widehat{\mathbb{Y}} |N,\lambda\rangle = \eta_N (-1)^{1/2-\lambda} |N,-\lambda\rangle
\end{align}
where $\widehat{\mathbb{Y}}$ is the parity transformation followed by a $180^\circ$ rotation
along the $y$-axis, and $\eta_N$ is internal parity, $\eta_N = 1$ for the nucleon and $\eta_N = -1$ for $N^*(1535)$.
Thus  $k^\perp \stackrel{\widehat{\mathbb{Y}}}{\mapsto} \bar k^\perp$,
$   |N,\up\rangle  \ \stackrel{\widehat{\mathbb{Y}}}{\mapsto} \ \eta_N |N,\down\rangle  \,,$
whereas for the quark states $q=u,d$
$  b_{q\uparrow}^{\dagger i}  | 0 \rangle \  \stackrel{\widehat{\mathbb{Y}}}{\mapsto} \ b_{q\downarrow}^{\dagger i}  | 0 \rangle\,,$
but
$  b_{q\downarrow}^{\dagger i}  | 0 \rangle \ \stackrel{\widehat{\mathbb{Y}}}{\mapsto} \ - b_{q\uparrow}^{\dagger i}  | 0 \rangle\,.$
Applying this transformation to the both sides of Eq.~(\ref{NucleonLCWF}) one obtains~\cite{Ji:2002xn}
\begin{widetext}
\begin{align}\label{NucleonLCWF1}
|N\down\rangle_{uud}^{\ell_z=0}&= -\eta_N\frac{\epsilon^{ijk}}{\sqrt{6}}\int\frac{[dx][dk_\perp]}{\sqrt{x_1 x_2 x_3}}
\Big[\psi_{N;1}^{(0)}(1,2,3) + i\epsilon^{\alpha\beta} \bar k^\perp_{1\alpha} \bar {k}^\perp_{2\beta}\psi_{N;2}^{(0)}(1,2,3)\Big]
b_{u\downarrow}^{\dagger i}(1) \Bigl(b_{u\uparrow}^{\dagger j}(2)b_{d\downarrow}^{\dagger k}(3)- b_{d\uparrow}^{\dagger j}(2)b_{u\downarrow}^{\dagger k}(3)
\Bigr)|0\rangle\,,
\notag\\
|N\down\rangle_{uud}^{\ell_z=-1}&= \eta_N\frac{\epsilon^{ijk}}{\sqrt{6}}\int \frac{[dx][dk_\perp]}{\sqrt{x_1 x_2 x_3}}
\Big[\bar k^\perp_1\psi_{N;1}^{(1)}(1,2,3) + \bar k^\perp_2 \psi_{N;2}^{(1)}(1,2,3)\Big]
\Bigl(b_{u\downarrow}^{\dagger i}(1) b_{u\uparrow}^{\dagger j}(2)b_{d\uparrow}^{\dagger k}(3)-
b_{d\downarrow}^{\dagger i}(1)b_{u\uparrow}^{\dagger j}(2)b_{u\uparrow}^{\dagger k}(3)
\Bigr)|0\rangle\,,
\notag\\
|N\down\rangle_{uud}^{\ell_z=1}&=
\eta_N \frac{\epsilon^{ijk}}{\sqrt{6}}\int \frac{[dx][dk_\perp]}{\sqrt{x_1 x_2 x_3}}\Big[k^\perp_1 \psi_{N}^{(-1)}(1,2,3)\Big]
b_{u\downarrow}^{\dagger i}(1) \Bigl(b_{u\downarrow}^{\dagger j}(2)b_{d\downarrow}^{\dagger k}(3)-b_{d\downarrow}^{\dagger j}(2)b_{u\downarrow}^{\dagger k}(3)\Bigr)|0\rangle\,,
\notag\\
|N\down\rangle_{uud}^{\ell_z=-2}&=
-\eta_N \frac{\epsilon^{ijk}}{\sqrt{6}}\int\frac{[dx][dk_\perp]}{\sqrt{x_1 x_2 x_3}}
\Big[\bar k^\perp_1 \bar k^\perp_3  \psi_{N}^{(2)}(1,2,3)\Big]
\Bigl(b_{u\uparrow}^{\dagger i}(1) b_{u\uparrow}^{\dagger j}(2)b_{d\uparrow}^{\dagger k}(3)-
b_{u\uparrow}^{\dagger i}(1)b_{d\uparrow}^{\dagger j}(2)b_{u\uparrow}^{\dagger k}(3)\Bigr)|0\rangle\,.
\end{align}
\end{widetext}
so that, e.g., for the $\ell_z=0$ states
\begin{align}
   \psi_{N;1}^{(0)}(1,2,3)\Big|_{N\down} = - \eta_N  \psi_{N;1}^{(0)}(1,2,3)\Big|_{N\up}
\end{align}

A Lorentz-covariant definition of the DAs of negative parity resonances involves some freedom.
It is convenient to choose the definition is such a way that the coefficient functions
in the OPE of currents (\ref{CF}) are the same for states of both parities, and also
the relations between different DAs imposed by QCD equations of motion remain the same.
As noticed in Ref.~(\cite{Braun:2009jy}), this can be achieved using
invariant decomposition of the $N^*(1535)$ matrix element in terms of
the $\gamma_5$-rotated quark fields
\begin{widetext}
\begin{eqnarray}
 4 (\gamma_5)_{\alpha\alpha'}(\gamma_5)_{\beta\beta'}(\gamma_5)_{\gamma\gamma'}
\langle 0| \epsilon^{ijk} u_{\alpha'}^i(a_1 n) u_{\beta'}^j(a_2 n) d_{\gamma'}^k(a_3 n) |N^\ast(P,\lambda)\rangle
&=&
S^{N^\ast}_1 m_{N^\ast} C_{\alpha \beta} \left(\gamma_5 u_{N^{\ast}}^+\right)_\gamma + \ldots
\end{eqnarray}
where the expression on the right hand side is the same as in Eq.~(\ref{defdisamp}) with obvious
replacements $m_N \to m_{N^\ast}$ etc. Projecting out the $\gamma_5$ matrices we obtain
\begin{eqnarray}
\label{defdisampstar}
\lefteqn{ 4 \langle 0| \epsilon^{ijk} u_\alpha^i(a_1 n) u_\beta^j(a_2 n) d_\gamma^k(a_3 n)
|N^\ast(P,\lambda)\rangle =}
\nonumber \\
&=&
S^{N^\ast}_1 m_{N^\ast} C_{\alpha \beta} \left(u_{N^\ast}^+\right)_\gamma +
S^{N^\ast}_2 m_{N^\ast} C_{\alpha \beta} \left(u_{N^\ast}^-\right)_\gamma +
P^{N^\ast}_1 m_{N^\ast} \left(\gamma_5 C\right)_{\alpha \beta} (\gamma_5u^+_{N^\ast})_\gamma +
P^{N^\ast}_2 m_{N^\ast} \left(\gamma_5 C \right)_{\alpha \beta} (\gamma_5u_{N^\ast}^-)_\gamma
\nonumber \\
&&
-V^{N^\ast}_1  \left(\!\not\!{p}C \right)_{\alpha \beta} \left(u_{N^\ast}^+\right)_\gamma
-V^{N^\ast}_2  \left(\!\not\!{p}C \right)_{\alpha \beta} \left(u_{N^\ast}^-\right)_\gamma
+ \frac12 V^{N^\ast}_3 m_{N^\ast}  \left(\gamma_\perp C \right)_{\alpha \beta}\left( \gamma^{\perp}
 u_{N^\ast}^+\right)_\gamma
\nonumber \\
&& +
\frac12 V^{N^\ast}_4 m_{N^\ast}  \left(\gamma_\perp C \right)_{\alpha \beta}\left( \gamma^{\perp}
u_{N^\ast}^-\right)_\gamma
-V^{N^\ast}_5  \frac{m_{N^\ast}^2}{2 p n}
\left(\!\not\!{n}C \right)_{\alpha \beta} \left(u_{N^\ast}^+\right)_\gamma
-\frac{m_{N^\ast}^2}{2 pn} V^{N^\ast}_6  \left(\!\not\!{n}C \right)_{\alpha \beta} \left(u_{N^\ast}^-\right)_\gamma
\nonumber \\
&&
- A^{N^\ast}_1  \left(\!\not\!{p}\gamma_5 C \right)_{\alpha \beta} (\gamma_5u_{N^\ast}^+)_\gamma
- A^{N^\ast}_2  \left(\!\not\!{p}\gamma_5 C \right)_{\alpha \beta} (\gamma_5u_{N^\ast}^-)_\gamma +
\frac12 A^{N^\ast}_3 m_{N^\ast}  \left(\gamma_\perp \gamma_5 C \right)_{\alpha \beta}\left( \gamma^{\perp}
\gamma_5u_{N^\ast}^+\right)_\gamma
\nonumber \\
&&
+\frac12 A^{N^\ast}_4 m_{N^\ast}  \left(\gamma_\perp \gamma_5 C \right)_{\alpha \beta}\left( \gamma^{\perp}
\gamma_5 u_{N^\ast}^-\right)_\gamma
 - A^{N^\ast}_5  \frac{m_{N^\ast}^2}{2 p n}
\left(\!\not\!{n}\gamma_5 C \right)_{\alpha \beta} (\gamma_5u_{N^\ast}^+)_\gamma
- \frac{m_{N^\ast}^2}{2 pn}  A^{N^\ast}_6  \left(\!\not\!{n}\gamma_5 C \right)_{\alpha \beta} (\gamma_5u_{N^\ast}^-)_\gamma
\nonumber \\
&&
-T^{N^\ast}_1 \left(i \sigma_{\perp p} C\right)_{\alpha \beta} \left(\gamma^\perp u_{N^\ast}^+\right)_\gamma
-T^{N^\ast}_2 \left(i \sigma_{\perp\, p} C\right)_{\alpha \beta} \left(\gamma^\perp u_{N^\ast}^-\right)_\gamma +
T^{N^\ast}_3 \frac{m_{N^\ast}}{p n}
\left(i \sigma_{p\, n} C\right)_{\alpha \beta}
\left(u_{N^\ast}^+\right)_\gamma
\nonumber \\
&&
+ T^{N^\ast}_4 \frac{m_{N^\ast}}{p n}\left(i \sigma_{n\, p} C\right)_{\alpha \beta} \left(u_{N^\ast}^-\right)_\gamma
- T^{N^\ast}_5 \frac{m_{N^\ast}^2}{2 p n}  \left(i \sigma_{\perp\, n} C\right)_{\alpha \beta}
\left(\gamma^\perp u_{N^\ast}^+\right)_\gamma
- \frac{m_{N^\ast}^2}{2 pn}  T^{N^\ast}_6 \left(i \sigma_{\perp\, n} C\right)_{\alpha \beta}
\left(\gamma^\perp u_{N^\ast}^-\right)_\gamma
\nonumber \\
&& + \frac12 m_{N^\ast} T^{N^\ast}_7 \left(\sigma_{\perp\, \perp'} C\right)_{\alpha \beta}
\big(\sigma^{\perp\, \perp'} u_{N^\ast}^+\big)_\gamma +
\frac12 m_{N^\ast} T^{N^\ast}_8  \left(\sigma_{\perp\, \perp'} C\right)_{\alpha \beta}
\big(\sigma^{\perp\, \perp'} u_{N^\ast}^-\big)_\gamma\,,
\end{eqnarray}
\end{widetext}
This expression replaces the decomposition (\ref{defdisamp}) for the nucleon.
Note that there are some minus signs and in particular all three leading twist DAs $V_1$, $A_1$ and
$T_1$ are defined in our convention with a different sign as compared to the nucleon.
As a consequence in the definition of leading-twist DA in terms of the chiral quark fields
there is a minus sign as compared to~(\ref{NDA:twist3}),
\begin{align}
\label{NDA:twist3star}
\ide{2em}
\langle 0 | \epsilon^{ijk}\!
\left(u^{\up}_i(a_1 n) C \!\!\not\!{n} u^{\down}_j(a_2 n)\right)
\!\not\!{n} d^{\up}_k(a_3 n) |N^*(P)\rangle \nonumber\\
&= \frac12 f_{N^*}\, (pn)\slashed{n}\, u_{N^*}^{\up}(P)\!
\!\int\! [dx] \,e^{-i (pn) \sum x_i a_i}\,
\varphi_{N^*}(x_i)\,,
\end{align}
where, of course, $P^2=m_{N^*}^2$ and the
expressions for the invariant functions $V_1^{N^\ast},A_1^{N^\ast},T_1^{N^\ast}$ in terms of $\varphi_{N^*}$,
are the same as for the nucleon, Eq.~(\ref{VAT1}).\\[2mm]

The twist-four DAs also acquire some signs~\cite{Braun:2009jy}
\begin{align}
\label{Phi4star}
\ide{2em}
\langle 0 | \epsilon^{ijk}\!
\left(u^{\up}_i(a_1 n) C\slashed{n} u^{\down}_j(a_2 n)\right)
\!\slashed{p} d^{\up}_k(a_3 n) |N^*(P)\rangle \nonumber\\
=& \frac14
\,(pn)\, \slashed{p}\, u_{N^*}^{\up}(P)\! \!\int\! [dx]
\,e^{-i (pn) \sum x_i a_i}\,
\nonumber\\
&\times \left[f_{N^*}\Phi^{N^*,WW}_4(x_i)+\lambda^*_1\Phi^{N^*}_4(x_i)\right],
\end{align}
\begin{align}
\label{Psi4star}
\ide{2em}
\langle 0 | \epsilon^{ijk}\!
\left(u^{\up}_i(a_1 n) C \slashed{n}\gamma_{\perp}\slashed{p} u^{\down}_j(a_2 n)\right)
\gamma^{\perp}\slashed{n} d^{\up}_k(a_3 n) |N^*(P)\rangle \nonumber\\
=&
-\frac12
\, (pn)\! \not\!{n}\,m_{N^*} u_{N^*}^{\up}(P)\! \!\int\! [dx]
\,e^{-i (pn)\sum x_i a_i}\,
\nonumber\\
&\times \left[f_{N^*}\Psi^{N^*,WW}_4(x_i)-\lambda^*_1\Psi^{N^*}_4(x_i)\right],
\end{align}
\begin{align}
\label{Xi4star}
\ide{2em}
\langle 0 | \epsilon^{ijk}\!
\left(u^{\up}_i(a_1 n) C\slashed{p}\slashed{n} u^{\up}_j(a_2 n)\right)
\!\not\!{n} d^{\up}_k(a_3 n) |N^*(p)\rangle \nonumber\\
=& \frac{\lambda^*_2}{12}\, (pn) \slashed{n}\, m_{N^*} u_{N^*}^{\up}(P)\! \!\int\! [dx]
\,e^{-i (pn) \sum x_i a_i}\,\Xi^{N^*}_4(x_i) \,,
\end{align}
where $\Phi^{N^*,WW}_4(x_i)$ and $\Psi^{N^*,WW}_4(x_i)$ are given by the
same expressions in terms of the expansion of the
leading-twist DA $\varphi_{N^*}(x_i)$ as for the nucleon.

\begin{widetext}
The price to pay for universality of correlation functions for positive and negative parities
is that the relations between DAs and light-front wave functions in this convention
acquire some signs as well,
\begin{eqnarray}
 f_{N^\ast} \varphi_{N^\ast}(x_1,x_2,x_3) &=& +4\sqrt{6} \int [dk_\perp] \,\psi_{N^\ast;1}^{(0)}(1,2,3)\,,
\nonumber\\
{}[\lambda^{N^\ast}_1\Phi^{N^\ast}_4+ f_{N}\Phi^{N^\ast,WW}_4](x_2,x_1,x_3)
&=&
+8\sqrt{6}\int \frac{[dk_\perp]}{x_3 m_{N^\ast}} k^\perp_3\cdot\Big[\bar k^\perp_1\psi_{N^\ast;1}^{(1)} + \bar k^\perp_2 \psi_{N^\ast;2}^{(1)}\Big](1,2,3)\,,
\nonumber\\
{}[\lambda^{N^\ast}_1\Psi^{N^\ast}_4 - f_{N}\Psi^{N^\ast,WW}_4](x_1,x_2,x_3)
&=&
-8\sqrt{6} \int \frac{[dk_\perp]}{x_2 m_{N^\ast}} \bar k^\perp_2\cdot\Big[k^\perp_1\psi_{N^\ast;1}^{(1)} + k^\perp_2 \psi_{N^\ast;2}^{(1)}\Big](1,2,3)\,,
\nonumber\\
\lambda^{N^\ast}_2 \Xi^{N^\ast}_4(x_1,x_2,x_3)
&=&
-24\sqrt{6} \int \frac{[dk_\perp]}{x_1 m_{N^\ast}} k^\perp_1\cdot \Big[\bar k^\perp_1\Big(\psi_{N^\ast}^{(-1)}(1,3,2)-\psi_{N^\ast}^{(-1)}(1,2,3)\Big)
\nonumber\\&&{}\hspace*{2.99cm}
+ \bar k^\perp_2\Big(\psi_{N^\ast}^{(-1)}(2,3,1)-\psi_{N^\ast}^{(-1)}(2,1,3)\Big)\Big]\,,
\end{eqnarray}
\end{widetext}
that have to be taken into account for the interpretation of the results.

Parametrization of the DAs of the resonances can be overtaken from that for the nucleon.
The leading-twist DA $\varphi_{N^\ast}(x_i,\mu)$ can be expanded in the set of orthogonal polynomials
$\mathcal{P}_{nk}(x_i)$
\begin{eqnarray}
 && \varphi_N(x_i,\mu) ~=~ 120 x_1 x_2 x_3 \sum_{n=0}^\infty\sum_{k=0}^n \varphi_{nk}(\mu) \mathcal{P}_{nk}(x_i)\,,
\nonumber\\
 && \int [dx]\, x_1 x_2 x_3 \mathcal{P}_{nk}(x_i)\mathcal{P}_{n'k'}(x_i) \propto \delta_{nn'}\delta_{kk'}\,,
\label{expand-varphi}
\end{eqnarray}
such that the coefficients are renormalized multiplicatively to one-loop accuracy,
\begin{eqnarray}
  f_{N^\ast}(\mu) &=& f_{N^\ast}(\mu_0) \left(\frac{\alpha_s(\mu)}{\alpha_s(\mu_0)}\right)^{2/(3\beta_0)}\,,
\nonumber\\
  \varphi_{nk}(\mu) &=& \varphi_{nk}(\mu_0)\left(\frac{\alpha_s(\mu)}{\alpha_s(\mu_0)}\right)^{\gamma_{nk}/\beta_0}.
\end{eqnarray}
Here $\beta_0 = 11-\frac23 n_f$ is the first coefficient of the 
QCD $\beta$-function and $\gamma_{nk}$ are the anomalous dimensions. The double sum in Eq.~(\ref{expand-varphi})
goes over a complete set of orthogonal polynomials $\mathcal{P}_{nk}(x_i)$, $k=0,\ldots,n$, of degree $n$:
\begin{eqnarray}
\mathcal{P}_{00} &= & 1\,,
\nonumber\\
\mathcal{P}_{10} &= &21(x_1-x_3)\,,
\qquad
\mathcal{P}_{11} = 7(x_1-2x_2 + x_3)\,,
\nonumber\\
\mathcal{P}_{20} &= & \frac{63}{10}[3 (x_1-x_3)^2 -3x_2(x_1+x_3)+ 2x_2^2]\,,
\nonumber\\
\mathcal{P}_{21} & = &\frac{63}{2}(x_1-3x_2 + x_3)(x_1-x_3)\,,
\nonumber\\
\mathcal{P}_{22} & = & \frac{9}{5} [x_1^2\!+\!9 x_2(x_1\!+\!x_3)\!-\!12x_1x_3\!-\!6x_2^2\!+\!x_3^2]
\label{lowest-P}
\end{eqnarray}
etc., and the corresponding anomalous dimensions are
\begin{align}
  \gamma_{00} =& 0\,, \qquad\quad\! \gamma_{10} = \frac{20}{9}\,,\qquad \gamma_{11} = \frac83\,,
\notag\\
 \gamma_{20} =& \frac{32}{9}\,,\qquad \gamma_{21} = \frac{40}{9}\,,\qquad \gamma_{22} =\frac{14}{3}\,.
\end{align}
The normalization condition (\ref{norm}) implies that $\varphi_{00}=1$.
In the main text we refer to the coefficients $\varphi_{nk}(\mu_0)$ with $n=1,2,\ldots $,  as shape parameters.
The set of these coefficients together with the normalization constant $f_N(\mu_0)$ at a reference
scale $\mu_0$ specifies the momentum fraction distribution of valence quarks on the nucleon.
They are related to matrix elements of local gauge-invariant
three-quark operators and can be calculated, e.g., on the lattice~\cite{Braun:2009jy,Braun:2014wpa}.

The twist-four DAs can be parameterized as~\cite{Braun:2008ia}
\begin{eqnarray}
 \Phi^{N^\ast}_4(x_i,\mu) &=& 24 x_1 x_2 \Big\{1 + \eta_{10}(\mu) \mathcal{R}_{10}(x_3,x_1,x_2)
\nonumber\\&&{} - \eta_{11}(\mu) \mathcal{R}_{11}(x_3,x_1,x_2)\Big\},
\nonumber\\
 \Psi^{N^\ast}_4(x_i,\mu) &=& 24 x_1 x_3 \Big\{1 + \eta_{10}(\mu) \mathcal{R}_{10}(x_2,x_3,x_1)
\nonumber\\&&{} + \eta_{11}(\mu) \mathcal{R}_{11}(x_2,x_3,x_1)\Big\},
\nonumber\\
 \Xi^{N^\ast}_4(x_i,\mu) &=&  24 x_2 x_3 \Big\{1 + \frac{9}{4}\xi_{10}(\mu) \mathcal{R}_{10}(x_1,x_3,x_2)\Big\},
\nonumber\\
\label{PhiPsi4}
\end{eqnarray}
where
\begin{eqnarray}
 \mathcal{R}_{10}(x_1,x_2,x_3) &=& 4\left(x_1+x_2-\frac32 x_3\right)\,,
\nonumber\\
 \mathcal{R}_{11}(x_1,x_2,x_3) &=& \frac{20}{3}\left(x_1-x_2+\frac12 x_3\right)\,
\end{eqnarray}
and $\eta_{10}(\mu)$, $\eta_{11}(\mu)$, $\xi_{10}(\mu)$ are the  new shape parameters.
The corresponding one-loop anomalous dimensions are~\cite{Braun:2008ia}
\begin{equation}
\gamma_{10}^{(\eta)} =\frac{20}{9}\,,\qquad \gamma_{11}^{(\eta)} = 4\,,\qquad \gamma_{10}^{(\xi)} = \frac{10}{3}\,.
\end{equation}
For the twist-five DAs we take into account contributions of geometric twist-three and twist-four operators
as explained in Ref.~\cite{Anikin:2013aka}.

Note that the asymptotic DAs (at very large scales) for the
nucleon and the resonances are the same:
\begin{align}
& \varphi^{\rm as}(x_i) = 120 x_1 x_2 x_3\,,\quad \Phi^{\rm as}_4(x_i) =  24
x_1 x_2\,,\quad
\nonumber\\
& \Phi^{WW,{\rm as}}_4(x_i)= 24 x_1 x_2 (1+\frac{2}{3}(1-5x_3))\,,
\nonumber\\
& \Psi^{WW,{\rm as}}_4(x_i)= 24 x_1 x_3(1+\frac{2}{3}(1-5x_2))\,,
\nonumber\\
& \Xi_4(x_i) = 24 x_2 x_3\,,\quad \Psi^{\rm as}_4(x_i) = 24 x_1 x_3\,.
\end{align}

For completeness we also give here the definitions of the
normalization constants in terms of matrix elements of local
three-quark operators:
\begin{align} \label{eq:nstarnormconst}
& \langle 0 |\epsilon^{ijk}\!
\left(u_i C\slashed{n} u_j\right)\!(0)
\gamma_5\slashed{n} d_k(0) |N^*\!(P)\rangle
\nonumber \\
& \hspace*{4cm}
 = f_{N^*} (pn) \gamma_5 \slashed{n}\, u_{N^*}\!(P),
\nonumber\\
& \langle 0 | \epsilon^{ijk}\!
\left(u_i C\gamma_{\mu} u_j\right)\!(0)
\gamma_5\gamma^{\mu} d_k(0) |N^*\!(p)\rangle
\nonumber \\
& \hspace*{4.0cm} = \lambda_1^{N^*} m_{N^*} \gamma_5 u_{N^*}\!(P),
\nonumber
\\
& \langle 0 | \epsilon^{ijk}\!
\left(u_i C\sigma_{\mu\nu} u_j\right)\!(0)
\gamma_5\sigma^{\mu \nu} d_k(0) |N^*\!(P)\rangle
\nonumber \\
& \hspace*{4.0cm} = \lambda_2^{N^*} m_{N^*} \gamma_5 u_{N^*}\!(P).
\end{align}

\vspace*{0.2cm}

\section{Parametrization of coefficient functions }

For convenience we provide a simple parametrization for the coefficient
functions $f^{nk}_{1,2}$, $g^{nk}_{1,2}$ appearing in (\ref{G1param}), (\ref{G2param}),
for the range $2 < Q^2 < 12~\text{GeV}^2$:
\begin{align}
 f^{nk}_{1,2} (Q^2) &= D(Q^2) \sum\limits_{p=0}^{4} b^{nk}_{p;1,2} \left(\frac{m_{N^\ast}^2}{Q^2}\right)^p,
\notag\\
 g^{nk}_{1,2} (Q^2) &= D(Q^2) \sum\limits_{p=0}^{4} a^{nk}_{p;1,2} \left(\frac{m_{N^\ast}^2}{Q^2}\right)^p,
\end{align}
where $D(Q^2)$ is the dipole form factor (\ref{dipole}).
The coefficients $a^{nk}_{p;1}$, $a^{nk}_{p;2}$, $b^{nk}_{p;1}$ and $ b^{nk}_{p;2}$ are collected in Table~\ref{tab:cf2}.

\begin{table*}[ht]
\renewcommand{\arraystretch}{1.2}
\begin{center}
\begin{tabular}{@{}l|l|l|l|l|l|l|l|l|l@{}} \hline\hline
              &  $a_{p;1}^{00}$   &  $a_{p;1}^{10}$   &  $a_{p;1}^{11}$  & $b_{p;1}^{00}$ & $b_{p;1}^{10}$ & $b_{p;1}^{11}$ & $b_{p;1}^{20}$ & $b_{p;1}^{21}$ & $b_{p;1}^{22}$  \\ \hline
$ p=0 $       & 0.0147491 & 0.251939 & 0.0256977 & 0.00716919 & 0.192078 & -0.0271761 & -0.340351 & -0.653521 & 0.00864077 \\\hline
$ p=1 $       & 0.773867 & -6.0864 & 0.646869 & -0.307557 & -1.94975 & -0.706607 & 4.61356 & 3.7677 & 0.112153 \\\hline
$ p=2 $       & -0.18913 & 5.62993 & 0.0535879 & -0.242484 & 0.246661 & 5.43478 & -1.81907 & -2.25019 & -0.258147 \\\hline
$ p=3 $       & 0. & -1.88333 & 0. & 0. & 0. & -5.79459 & -2.75613 & 0. & 0.0954517 \\\hline
$ p=4 $       & 0. & 0. & 0. & 0. & 0. & 1.99163 & 1.67516 & 0. & 0. \\\hline
\end{tabular}
\vskip0.3cm
\begin{tabular}{@{}l|l|l|l|l|l|l|l|l|l@{}} \hline\hline
              &  $a_{p;2}^{00}$   &  $a_{p;2}^{10}$   &  $a_{p;2}^{11}$  & $b_{p;2}^{00}$ & $b_{p;2}^{10}$ & $b_{p;2}^{11}$ & $b_{p;2}^{20}$ & $b_{p;2}^{21}$ & $b_{p;2}^{22}$  \\ \hline
$ p=0 $       & 0.0469231 & -0.365146 & -0.0498471 & 0.13253 & 0.210365 & 0.763116 & -0.898009 & 0.978028 & -0.408284 \\\hline
$ p=1 $       & -1.35098 & 10.1647 & 1.78846 & -1.41541 & 0.1675 & -2.84988 & 23.7579 & 19.1668 & 7.37946 \\\hline
$ p=2 $       & 1.30792 & -16.9382 & -3.92016 & 0.522203 & -3.43005 & 1.08085 & -50.8692 & -47.2691 & -14.8858 \\\hline
$ p=3 $       & -0.450538 & 12.7283 & 3.3271 & 0. & 2.01318 & 0. & 46.6823 & 43.7722 & 12.2449 \\\hline
$ p=4 $       & 0. & -3.67258 & -1.03533 & 0. & 0. & 0. & -16.0002 & -14.5341 & -3.76771 \\\hline
\end{tabular}
\end{center}
\caption[]{\sf Coefficient functions in the LCSRs for $N^\ast(1535)$ production}
\label{tab:cf2}
\renewcommand{\arraystretch}{1.0}
\end{table*}




\end{document}